# VISIONS IN THEORETICAL COMPUTER SCIENCE

## A Report on the TCS Visioning Workshop 2020

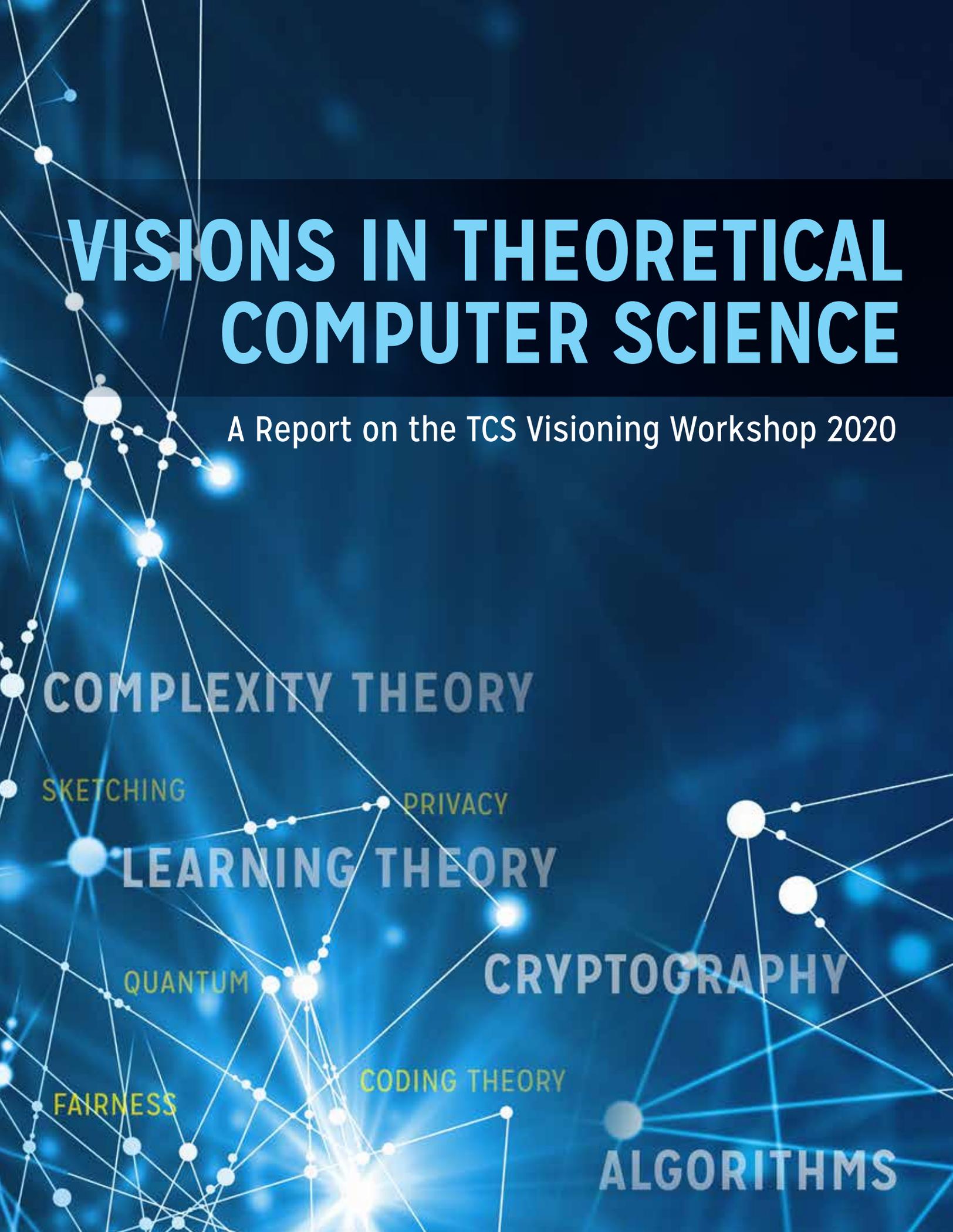

The material is based upon work supported by the National Science Foundation under Grants No. 1136993 and No. 1734706. Any opinions, findings, and conclusions or recommendations expressed in this material are those of the authors and do not necessarily reflect the views of the National Science Foundation.

# VISIONS IN THEORETICAL COMPUTER SCIENCE

## A REPORT ON THE TCS VISIONING WORKSHOP 2020

**Edited by:**

**Shuchi Chawla (University of Wisconsin-Madison)**
**Jelani Nelson (University of California, Berkeley)**
**Chris Umans (California Institute of Technology)**
**David Woodruff (Carnegie Mellon University)**

# Table of Contents

# Foreword

Theoretical computer science (TCS) is a subdiscipline of computer science that studies the mathematical foundations of computational and algorithmic processes and interactions. Work in this field is often recognized by its emphasis on mathematical technique and rigor. At the heart of the field are questions surrounding the nature of computation: What does it mean to compute? What is computable? And how efficiently? These lines of inquiry applied to different models and settings of computation have resulted in new disciplines within and beyond computer science including learning theory, cryptography and security, quantum computing, economics and computation, computational biology and computational geometry. In recent decades, the field expanded its reach into many varied topics at the forefront of technology, physical sciences and social sciences. The infiltration into different areas helped bring mathematical rigor and algorithmic thinking to areas such as privacy, fairness, and data science. This breadth of application areas is interconnected through a technical core of computational complexity and algorithm design.

This document describes some recent and current directions of research within TCS, how these directions contribute to the betterment of the discipline and the betterment of society and some of the technical challenges that lie ahead. A primary goal of this document is to convince the reader that strength in theoretical computer science is vital to the health of the computer science field as a whole, as well as, the continuation of technological progress in the 21st century.

Theoretical Computer Science impacts computing and society by identifying key issues in new areas and framing them in ways that drive development. In fact much of the history of Computer Science, as viewed through the lens of Turing Award citations, is filled with examples of major fields that were pioneered by TCS researchers: cryptography (Adleman, Rivest, Shamir, Micali, Goldwasser); the modern theory of algorithms and computational complexity (Cook, Karp, Hopcroft, Tarjan, Hartmanis, Stearns, Blum, Yao); the foundations of machine learning (Valiant); and distributed systems (Lamport, Liskov). More recently, TCS has played a central role in the creation of the fields of quantum computation, algorithmic economics, algorithmic privacy and algorithmic fairness.

A unique feature of Theoretical Computer Science is its ability to discern computation and algorithms in settings beyond Computer Science proper. Examples include neuroscience, where one models systems within the brain and nervous system as computation by networks of discrete elements (for a seminal example of this perspective, see Valiant's "Circuits of the Mind"); systems biology, where networks of interacting genes or proteins are understood via computational models; economics, where self-interested interacting entities can be naturally seen to be performing a distributed computation; and physics, where the fundamental notions of information and computation, and insights from TCS, are deeply entwined in the current frontiers of research in gravity and quantum mechanics. Breakthrough results in pure mathematics and statistics by TCS researchers are becoming increasingly common (for example, the refutation of the Connes Embedding Conjecture, and the proof of the Kadison-Singer Conjecture).

Key technologies that grew out of Theoretical Computer Science have had a major impact in industry. Examples include the discovery of the principles that led to Google's PageRank algorithm; the development of Consistent Hashing, which in large part spawned Akamai; various key innovations in coding theory such as rateless expander codes for fast streaming, polar codes as part of the 5G standard, and local reconstruction codes for cloud storage; fast, dynamic algorithms that form the basis of navigation systems such as Google Maps and Waze, and other applications; quantum computing; cryptographic innovations such as public-key encryption and multiparty computation that form the foundation of a secure Internet; and the cryptographic underpinnings of cryptocurrencies and blockchain technology.

Despite this wide range of application areas and intersecting scientific disciplines, the field of TCS has thrived on the basis of a strong identity, a supportive community, and a unifying scientific aesthetic. The field is defined by a well-developed common language and approach that exposes previously unknown connections between disparate topics, thereby driving progress.

This document presents the case for robust support of foundational work in TCS, structured so as to allow unfettered exploration guided by the opportunities and needs of a rapidly changing and dynamic field, and thereby bringing about the greatest possible impact to society. This model has been extraordinarily successful in the past.

## A STRONG THEORETICAL COMPUTER SCIENCE FOUNDATION

The core of TCS encompasses some of the deepest and most fascinating questions in Computer Science and Mathematics, including the P vs. NP problem and many other fundamental questions about the possibilities and limitations of algorithms and computation. But it is also teeming with more modern problems that arise from an expansive view of computation as a language for understanding systems in many contexts. TCS has a unique way of framing these problems in terms of tradeoffs between computational resources; for example, between optimality and computational efficiency, online versus offline computation, centralized versus distributed equilibria, or degree of interaction versus security. Important fields such as streaming algorithms, rigorous cryptography, and algorithmic privacy were the direct outgrowth of this "TCS approach". Core TCS research focuses on a computational **problem** as the central object, rather than the development of prevailing algorithms and methods for a given problem. This viewpoint drives innovation and has led to unconventional approaches and solutions, often by identifying previously unknown connections between fields.

Importantly, these developments and many others arise from a **strongly supported core of TCS researchers** adept at identifying important problems from a broad range of settings, applying the tools of TCS, pulling in (and in some cases developing) sophisticated mathematics, and following the research wherever it leads.

Sustained and predictable investment in core TCS, supporting the best ideas wherever they arise, is key to continued innovation and success in the coming decade. While it is difficult to predict which advances will have the widest impact, past experience shows that investing in core TCS produces profound returns.

## HOW THIS DOCUMENT CAME ABOUT

Every ten years or so the TCS community attends visioning workshops to discuss the challenges and recent accomplishments in the TCS field. The workshops and the outputs they produce are meant both as a reflection for the TCS community and as guiding principles for interested investment partners. Concretely, the workshop output consists of a number of nuggets, each summarizing a particular point, that are synthesized in the form of a white paper and illustrated with graphics/slides produced by a professional graphic designer.

The second TCS Visioning Workshop was organized by the SIGACT Committee for the Advancement of Theoretical Computer Science and took place during the week of July 20, 2020. Despite the conference being virtual, there were over 76 participants, mostly from the United States, but also a few from Europe and Asia who were able to attend due to the online format.

The workshop included a very diverse set of participants including people from industry and from fields other than TCS. Represented sub areas included: Data-Driven Algorithms, Coding Theory and Communication, Complexity Theory, Optimization, Cryptography, Foundations of Machine Learning and Data Science, Sublinear Algorithms, Distributed Computing, Economics and Computer Science, Fairness and Social Good, Privacy, and Quantum Computing.

Workshop participants were divided into categories as reflected in the sections of this report: (1) models of computation; (2) foundations of data science; (3) cryptography; and (4) using theoretical computer science for other domains. Each group participated in a series of discussions that produced the nuggets below.

# Models of Computation

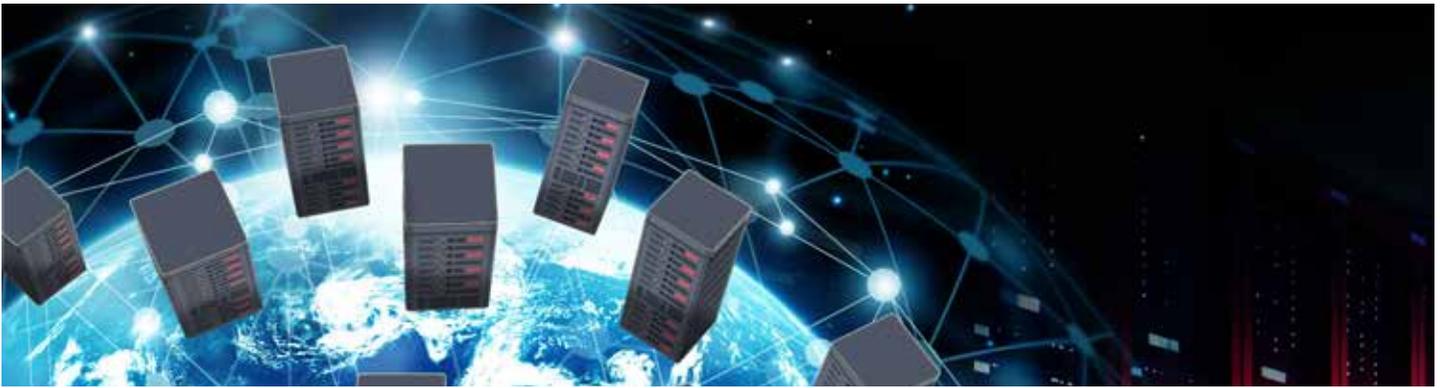

## COMPUTATIONAL COMPLEXITY

Complexity theory tries to classify computational problems according to the resources they use and then relate the classes to one another. The process often involves reductions of the form: any algorithm for solving Problem A can be used to solve Problem B, and thus, the resources required of an algorithm for solving Problem A must be, at least, those required of any algorithm for solving Problem B. Using such reductions, celebrated work has designed a universally optimal algorithm for a wide range of discrete optimization problems. While traditional complexity theory focuses on separating "easy" from "hard" problems, a recent trend looked at more fine-grained measures, relating problems to each other so that we can understand exactly how easy or how hard such problems are. Yet another explanation complexity theory has provided is how hard problems can be used as a source of randomness, that is, a sequence of bits may look random to an algorithm with limited resources, but non-random to more powerful algorithms.

## A Universal Algorithm for Discrete Optimization

*Automating algorithm design for finding the best solution.*

A wide range of societal, economic, and computational challenges can be viewed as discrete optimization tasks. These tasks include finding the best way to allocate constrained resources like trucks to shipments, creating schedules for airplanes, choosing the best routes for delivery, or finding the best representation of data for machine learning.

Research over the last two decades has shown that a single, simply-tunable (universal) algorithm – semidefinite programming based on sum-of-squares proofs (also known as the SOS algorithm) – is the most successful algorithm for all of these problems. Moreover, breakthrough work in complexity has indicated that the SOS algorithm is not only a good method, but is the best possible method among any that are currently and efficiently implementable.

## Unifying Modern Computational Challenges

*Fine-grained complexity connects disparate problems together.*

What do the following have in common: finding the differences between two documents, searching for pictures in your gallery resembling the one you just took, and uncovering small groups of mutual friends? For all of these search tasks, computer scientists know quick and inexpensive methods for solving them on moderately-sized examples. However, the same methods are challenged on larger examples, arising in the era of big data. When the "documents" are enormously-long human DNA strands, the galleries have billions of images, and the friends are on social networks with billions of users, these formerly-quick methods can become rather slow, consuming massive energy, memory, and other resources. The theory of fine-grained complexity provides a precise understanding of how the best methods to solve these and many other problems will scale with the size of the input data.

A web of interconnections has been revealed, showing that the challenges for many different domains are deeply linked through a few "core challenge" problems. These core problems reveal common bottlenecks that must be overcome to achieve more efficient programs in a wide range of applications. This gives evidence that the "slow down" of these algorithms on larger inputs is not an artifact of certain programs we run, but a deeper phenomenon: a computational "law of nature" to which any program must conform. Beyond the scientific benefits of understanding how core problems relate to each other, a fine-grained understanding of computational difficulty could be used to design new tools in cybersecurity, and improve blockchain technology (the basis for bitcoin and other e-currencies), where it is crucial to know precisely how much work must be done to solve a problem.

## Harnessing Unpredictability

### The power of random choices in computation.

Randomized algorithms pervade both the theory and practice of computer science. In this instance randomness can be defined as a resource rather than randomness in the input data. In other words, the algorithm is allowed to toss coins and make decisions based on the results, but the input to the algorithm is arbitrary. Perhaps surprisingly, randomized algorithms are often much more efficient for solving tasks than algorithms that do not make random choices.

Beyond computer science, randomness is an enigmatic phenomenon that has long fascinated and frustrated philosophers, physicists, and mathematicians alike. Complexity theory pioneered the realization that randomness is not absolute, but relative to the way it is used. Using this realization, complexity theory discovered a deep duality between randomness and the existence of "hard problems" — problems that cannot be solved by any efficient program. Paradoxically, this means that hard problems are useful in designing efficient programs in other areas. This has led to breakthroughs in many fields, including modern cryptography, privacy, scientific simulation and cloud computing.

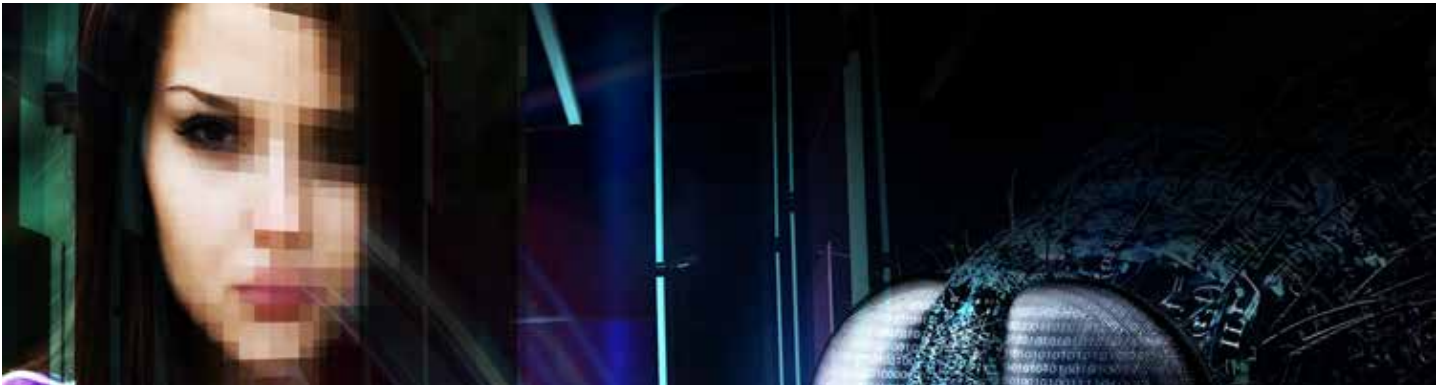

## SUBLINEAR ALGORITHMS

The large amount of data available in certain settings sometimes requires algorithms which do not have time to analyze all the data, and/or the memory to remember all the data. For example, the Large Hadron Collider generates one petabyte of data per second! We thus seek algorithms whose running time and/or memory usage are *sublinear* in the size of the data being processed. Such considerations have led to much work on algorithms which make just one pass over a stream of data ("streaming algorithms"), compressed data structures or data communication protocols with complexity far smaller than data size ("sketching"), and sublinear time algorithms which only look at a small fraction of data.

### Computing on Massive Datasets, made Easy by Sketching

*Matching photos and how the fruit fly's brain processes — what's in common?*

Data is abundant: over one billion photos are uploaded daily to just Google Photos. Analyzing such massive data can easily overwhelm available resources. For a typical photo matching task — grouping today's photos by their contents — a naive approach would perform more than quadrillions of photo comparisons, a task out of reach for computers. Sketching arose as a prominent TCS paradigm to perform such tasks by summarizing (sketching) the data to store only the essential information. This reduces the data size dramatically and makes the computational task easier. How does one sketch and use such sketches to enable efficient computation?

This question has inspired an exciting body of research in TCS that has impacted diverse applications including databases, machine learning, and signal processing (e.g., MRI). Sketching has enabled a new class of low-memory procedures that can process large amounts of real-time data, without storing all of it. Techniques in sketching have stimulated the development of a new field of randomized linear algebra that addresses key problems in scientific computing, network analysis, and machine learning. Strikingly, similar sketching methods have been discovered in nature, for instance, in how fruit flies process odors in their brain.

### Sublinear Network Analysis

*Studying big networks from small samples.*

Networks are ubiquitous. They model diverse phenomena including the transmission of information, social communication, and molecular interactions. We often cannot see the entire network; either it is too big, or too expensive to observe all of it. Yet, an understanding of these networks is key to many disciplines. Can we predict an epidemic spread by observing only a few social interactions? Can we learn the operation of spam networks by tracking only a few spammers? Can we understand the resilience of the nation's power grid by studying a few power stations?

Many questions follow this theme: a large network of importance, but the ability to only see a small portion of it. The study of sublinear algorithms provides the language and tools to rigorously describe and understand such problems. Given the explosive growth in network data, there is a compelling need to build a theory of sublinear network analysis that resonates with the various networks in the real world.

## Understanding the Brain

*Use ideas and methodologies from theory of algorithms to develop an algorithmic view of the brain.*

Understanding how the brain performs complex tasks is one of the biggest mysteries to elude scientists from many different fields including biology, neuroscience and computer science. When viewed from a computational standpoint, the challenge is identifying the key algorithmic ingredients and the right architecture to understand the brain conceptually; this algorithmic viewpoint allows us to think in an abstract way and disentangles the precise physiological details from the algorithmic abilities. TCS can study different tasks such as language understanding and propose algorithmic architectures for solving such tasks while backing them up with arguments or proofs that justify their efficacy. TCS can also help in modeling these tasks — for example, by providing a generative model for language.

There are several ideas in algorithms, in particular, data structures for similarity search, sketching, storage and indexing, and graph processing, all of which seem to be essential components of almost any major learning task that humans perform. For example, the way we access events from memory is almost always never exact, but tolerant to noise; thus we likely almost never store things in memory precisely, but in some concise "sketched" format. We perhaps store a graph of concepts or events based on how they are related or correlated. With its current set of algorithmic tools TCS can help us understand and propose how there may be special units in a "conceptual view" of the brain and how these units interact and work together to solve hard learning tasks. In addition to helping us understand the brain, this may also help us develop new ideas that may contribute to the field of algorithms and computer science.

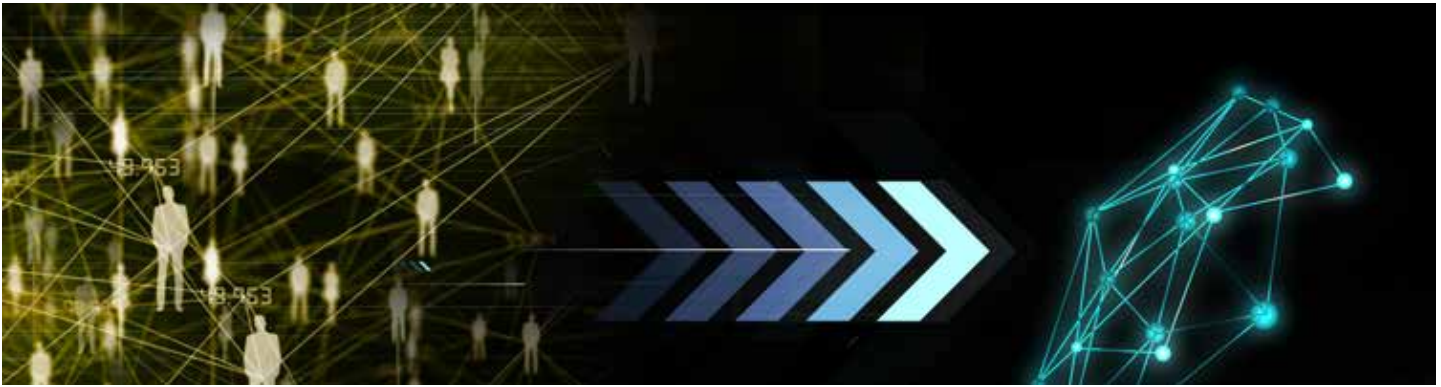

# CODING THEORY AND COMMUNICATION

Coding theory is the study of codes for specific applications. A code can be used for data compression, data transmission, and data storage. The goal is to design efficient and reliable codes to remove redundancy and to detect and correct errors. Given the recent explosion in the amount of data created by various applications, understanding how to design super-efficient codes that deal with frequently occurring errors on massive data sets is crucial. As inspiration for designing codes, we can look at the miraculous ability of various naturally occurring storage devices, such as our DNA. There are a number of challenges we must face if we want to make DNA storage a reality. Closely related to the ability to reliably store and encode information is the ability for two agents to reliably communicate and understand each other, which requires encoding information in a way that makes recognition and agreement between agents possible.

## DNA for Data Storage
### *Storing the world's data in a few grams of DNA.*

The storage density and longevity of DNA is a miracle of nature. Each cell in your body contains a tiny bit of DNA which encodes all of you. Not only does a single gram of DNA contain 215 million gigabytes of information, that information remains readable for several hundred thousands of years.

DNA storage is a tremendous emerging technology to deal with the exponentially increasing amounts of data we produce, which roughly doubles every two years. However, making DNA storage a reality requires a deep understanding of how to code for and deal with insertion and deletion errors that frequently occur when reading or writing DNA.

## Coding Theory for Computing
### *Getting the most out of imperfect hardware.*

Today, trillions of transistors, each being a few atoms wide, are performing massive computations and the computational demands continue to increase exponentially. Fundamental laws of physics show that accommodating such a growth will inevitably make the next-generation hardware prone to errors on every level of computation.

Classical error correcting codes, that have fueled the information revolution, are designed to protect data against errors. Scalable computation on future faulty hardware crucially requires new types of codes that can protect computation against errors. Such codes also play a key role for building quantum computers.

## A Theory of Explanations
### *What does it mean to explain "why" a system made a decision? What does it mean to "accept" an explanation?*

Theoretical computer science has an excellent record of formalizing social constructs, from privacy-preserving data analysis (differential privacy) to protocols for convincing a recipient of the veracity of a mathematical statement (interactive proofs, zero-knowledge) or verifying that a message did indeed originate with a claimed sender (digital signature). As algorithms reach ever more deeply, and more consequently, into our lives, there is a sense that their decisions should be "explainable", or "interpretable". Regulation may even require this ("There is no single, neat statutory provision labelled the 'right to explanatio' in Europe's new General Data Protection Regulation (GDPR). But nor is such a right illusory." — Selbst and Powles). What is the mathematical meaning of an explanation? Is an explanation more valuable if it is accepted by a party with greater resources, such as domain knowledge and computational ability, than if it is accepted by a party with limited resources? Are explanations unique, or can ex post justifications obscure the real reasons for a decision? Two natural starting points to answering these questions are the insights and tools in interactive proof systems and the work on goal-oriented universal semantic communication. These areas work to address the question of "how can two intelligent entities (or agents), that have no prior knowledge of each other, communicate to reach a state of understanding?"

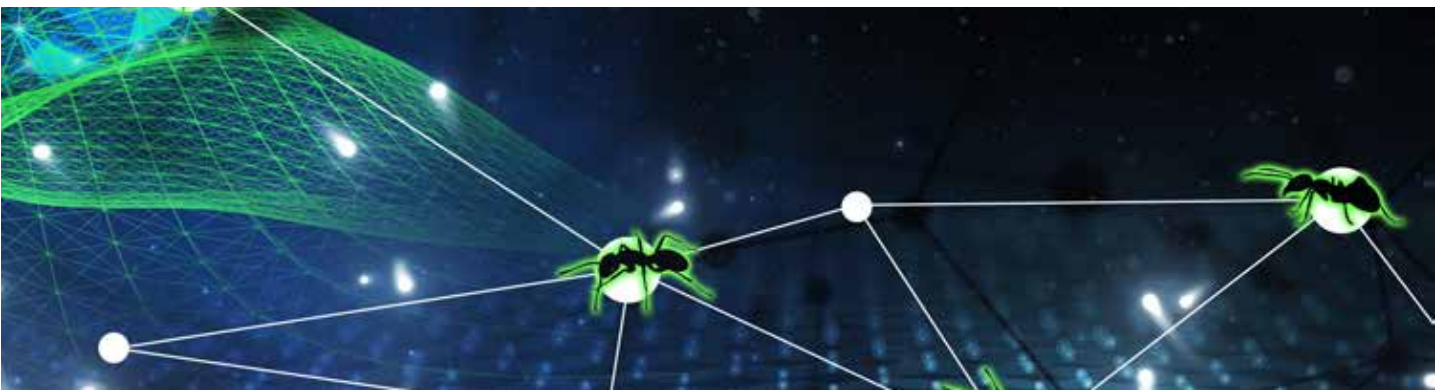

# DISTRIBUTED COMPUTING

The quantity of data we, as a society, generate and use on a daily basis is massive. It goes far beyond our capacity to store and process on any one machine. Furthermore, the data is shared by many users and accessed from many locations. A key research achievement of distributed computing has been distributed cloud storage systems that enable the big data technologies that are generating so much excitement today. These systems rely on fast algorithms for accessing shared data concurrently while updates are being performed. In addition, they rely on algorithms for maintaining consistent data efficiently over large distances. These algorithms were developed, in part, by unforeseen connections between biological distributed mechanisms and other distributed scenarios occurring in nature. Further understanding this connection will help expand our current technologies. This is especially important given the rapid growth of data, with scalability as a natural concern.

## Biological Distributed Algorithms

*Using algorithms to understand nature, and nature to inspire algorithms.*

The rapidly expanding field of biological distributed algorithms applies tools from the formal study of distributed computing systems to better understand distributed behavior in nature, and in turn uses these insights to inspire new ways to build computing systems. Results in recent years have revealed the surprising degree to which these two worlds are intertwined. For example, techniques adapted from the theoretical study of distributed systems helped uncover how ant colonies find food and allocate work, explain the eerie efficiency with which slime molds spread to locate food, identify simple strategies with which fireflies might desynchronize their flashes, unlock the capabilities of chemical reaction networks, and produce progress on open questions concerning how neurons in the human brain solve fundamental problems. Moving in the other direction, the models and techniques developed to study fireflies led to ideas used in breakthroughs in classical distributed graph problems and the algorithmic treatment of human neurons is unlocking new ideas for digital machine learning. A challenge is to further strengthen and learn from these connections, in order to mutually advance distributed algorithms, as well as, our understanding of naturally occurring distributed systems.

## Scalable Network Algorithms

*Computing on data too large to store in one place.*

Data is exponentially growing and vastly exceeds what can be computed on in a single machine. As a result, modern systems consist of many tens of thousands of networked servers running distributed algorithms. Delays, limited bandwidth, and lack of parallelizability are bottlenecks that arise and fundamentally new distributed algorithms are key to achieving solving these efficiency problems. Novel models allow us to develop algorithms for these massively parallel computations.

At the same time, new programming paradigms for large-scale graph processing, like Google's Pregel or Facebook's Giraph, implement interfaces precisely matching classical distributed message-passing models. Message-passing algorithms developed for these models now fuel the collection and analysis of network data. Making these algorithms even more efficient by going beyond worst-case guarantees remains an ongoing fundamental challenge with wide-ranging promise for future systems.

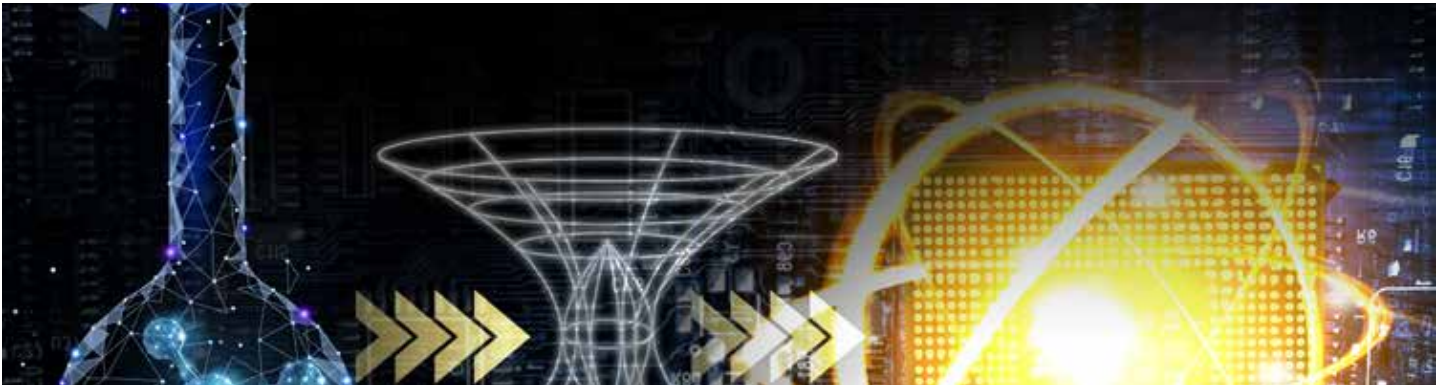

# QUANTUM COMPUTING

The ability of quantum computers to outperform classical computers is becoming a reality. How did quantum computers first beat classical ones? It took heroic experimental feats, ultracold temperatures, and a decade of work by theoretical computer scientists. This leads to natural questions, such as how to design even faster algorithms for quantum computers and how to verify their computations. Moreover, the ideas from quantum computing have been revolutionizing many areas of physics — from explaining new states of matter to the study of quantum gravity, wormholes, and Stephen Hawking's black hole information paradox. The next step is figuring out how to best harness this power to make further impacts on other scientific disciplines.

In the Fall of 2019, quantum computing achieved a historic milestone when Google announced the achievement of "quantum supremacy": that is, the first-ever use of a programmable quantum computer to perform some task that's believed to be much harder with any of the current conventional computers. Google's chip, called "Sycamore," used 53 superconducting qubits, cooled to a hundredth of a degree above absolute zero, to generate random 53-bit strings, in which the probabilities of some strings were enhanced by constructive quantum interference. The task was admittedly an artificial one — but Sycamore completed choreographing nine quadrillion amplitudes in quantum superposition in three minutes, while even conventional supercomputers with hundreds of thousands of cores are conjectured to need at least several days. This was first and foremost a remarkable feat of engineering — but John Martinis, who led the Google team responsible for it, has said that it wouldn't have happened had theoretical computer scientists not laid the groundwork. This experiment represents the moment at which quantum computing has finally entered the early "vacuum tube era," with actual experiments that will inform us about the possibility of quantum speedups for practical problems. Once you've built 50 or 60 noisy, programmably-coupled qubits, what should you have them do, to provide the most convincing possible speedup over a classical computer? How should you check the quantum computer's results? How confident can you be that the task in question really is classically hard? Since 2009, theoretical computer science has made a decisive contribution to answering all these questions, starting from abstract complexity theory and proceeding all the way to experimentation.

## Verifiable Quantum Computing

### *How can you check a quantum computer's calculation?*

Quantum computers are believed to have extravagant computational capabilities. While this means they have the potential to solve extremely complex problems, this power also makes it especially difficult to verify that they are behaving as intended. Over the last 15 years, theoretical computer scientists developed sophisticated methods to tackle the quantum verification problem. These methods leverage decades of theory research spanning complexity, cryptography, and physics. A couple of recent achievements exemplify the resulting significant theoretical and practical impact. The first example is the US National Institute of Standards and Technology's use of these methods in their "Randomness Beacon" to provide a source of certified quantum random numbers. Secondly, researchers demonstrated it is possible to use cryptography to check the calculations of a cloud quantum computer. And most recently, in early 2020 there was a breakthrough in quantum verification that solved the 46-year-old Connes' Embedding Conjecture from pure mathematics — spectacularly highlighting the far-reaching impact of research in verifiable quantum computing.

As quantum computing transitions from theory to practice in the upcoming years, questions of quantum verifiability take on increasing importance. For example, is there a way to verify the calculations of intermediate-scale, noisy quantum computers that are offered over the cloud? Can smaller quantum computers be used to check the results of larger quantum computers? Is there a tradeoff between privacy, efficiency, and verifiability of quantum computations? What types of quantum cryptosystems can be built using verification protocols? Theoretical computer science will continue to play a crucial role in finding the answer to these questions.

## A Quantum Computing Lens on Physics: From Exotic Matter to Black Holes

Over the last decade, ideas from quantum computing have been revolutionizing many areas of physics — from explaining new states of matter to the study of quantum gravity, wormholes, and Stephen Hawking's black hole information paradox.

One of the most important uses of quantum computers will be for performing large-scale simulations of quantum physics and quantum chemistry — for example, simulating large molecules for drug discovery or simulating exotic states of matter for materials design. In recent years, ideas from theoretical computer science such as random walks and gradient descent have led the way in developing faster and more practical algorithms for quantum simulations and they are on the verge of being run on real-world quantum hardware. However, many fundamental challenges remain. For example, can quantum computers developed in the near term be used to solve simulation problems that are intractable for classical computers? Can quantum computers aid us in designing exotic materials, such as high-temperature superconductors? Such research directions will require deep and sustained interdisciplinary collaborations with the physical sciences.

Even before it was thought to use quantum computing as a practical tool for materials science, ideas from quantum computing have become indispensable to fundamental physics. One of the central questions in this domain is how to combine the two great theories of 20th century physics: gravity and quantum mechanics. These ideas conflict most dramatically when considering black holes. Quantum mechanics predict that information is never created or destroyed, while Einstein's theory of gravity predicts that information thrown into a black hole is lost forever. Resolving this paradox is an important step towards establishing a theory of quantum gravity. Recently a promising approach to this puzzle was developed based on quantum error correcting codes. These codes were designed to protect quantum data against noise by using redundancy by spreading out information across many systems. This phenomenon turned out to be an effective way to describe the distorted space-time in the vicinity of a black hole. This revelation has led to the study of black holes, wormholes, and other puzzles of quantum gravity through the lens of quantum information — importing ideas from computer science including error correction, computational complexity and convex optimization.

Examples of real-world impact include:

1.  Roughly half of DOE supercomputer time is used for quantum simulation and quantum computers have the potential to both reduce the cost of these simulations and extend their scope.
2.  Molecular simulations are already of significant use in the pharmaceutical and chemical industries. Quantum computers could make existing simulations cheaper and enable new ones that would never have otherwise been possible.
3.  Improvements in chemistry and material science could mean stronger and lighter materials for cars, airplanes and other machines — better solar cells; improved superconductors; better catalysts for industrial processes.
4.  Recent advances in high-temperature superconductors are responsible for a promising recent approach to fusion power, but the underlying physics of these superconductors remains mysterious since they are too difficult for our existing computers to accurately model.

## New Frontiers in Quantum Algorithms

### *Which problems can be solved exponentially more quickly by quantum computers?*

Factoring numbers — say, factoring 21 into 3 x 7 — is one of the oldest problems in mathematics, and is the basis of most encryption used on the internet today. Factoring small numbers is relatively easy but when the numbers are thousands of digits long, our existing computers would take longer than the age of the universe to find the factors. However, two and a half decades ago, Peter Shor showed that a quantum computer could factor even large numbers quickly, setting off a scramble to find new codes to protect our communications and to understand what features of quantum mechanics enable this remarkable ability.

Since then, researchers have shown that quantum computers dramatically outperform their classical counterparts on a variety of other problems as well, including simulating physics and chemistry. In the future, we expect quantum computers to find even more applications. We list three of the most important potential applications below.

1. Machine learning and optimization have an enormous and still-growing range of applications in industry and science. Quantum computers can approach these problems in qualitatively different ways but our understanding of the power of these methods is still incomplete. Can quantum computers speed up machine learning and optimization or improve the quality of the answers?

2. Due to Shor's factoring algorithm, NIST is currently sifting through candidate cryptosystems to replace the factoring-based cryptosystems in use today. How can we tell if these "post-quantum cryptosystems" are actually secure against quantum computers? Trying to design quantum algorithms to break these systems is a crucial component of this area of research.

3. For now, and the foreseeable future we will have access to noisy intermediate-scale quantum (NISQ) computers, which are small quantum computers with hundreds or thousands of qubits rather than the many millions of qubits that a full-scale quantum computer requires. Can these NISQ machines give speed-ups, even though they are small?

# Foundations of Data Science

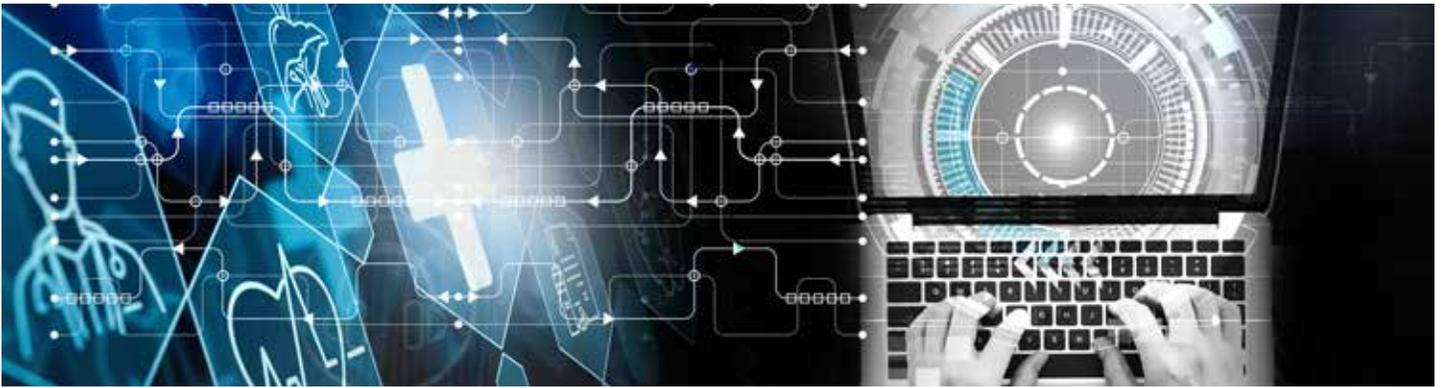

## DATA-DRIVEN ALGORITHM DESIGN

*Data Science* is a discipline focused on extracting knowledge from data using tools from mathematics, statistics, and computer science. Though the term was introduced more than 40 years ago, it only became widely used in the early to mid 2000s, with a surge in popularity in the last decade accompanied by several universities creating degree programs, institutes, and schools devoted to the subject. Theoretical computer science specifically, together with other fields, has played a major role in contributing to the foundations of the discipline, such as defining computational models, developing new efficient algorithms for them, and providing a computational lens on statistical questions.

Below we discuss avenues for further research in adaptive data analysis and protecting data from overuse. This includes using machine learning to augment algorithms while maintaining provable guarantees, self-improving algorithms that use feedback from data, and understanding special properties of datasets that are relevant in practice and that allow us to analyze algorithms beyond the worst case.

## Tools for Replicable Science

*Protecting the scientific value of data from overuse.*

Scientific and medical research is facing a "replication crisis". Many published results from these areas do not hold up when tested on new data. In other words, the original research identified a pattern that only occurred by chance in one dataset, rather than discovering an underlying truth in all datasets. This problem commonly arises when a dataset is over-used; as the saying goes, "if you torture the data long enough, it will confess to anything."

How can we design systems for data analysis that allow scientists to re-use the data across many studies and draw valid conclusions? The greatest challenge is that the design of later studies can depend on the results of earlier studies, making the usual statistical tools incorrect. This leads to false "discoveries" due to overconfidence in results' significance. The problem stems from the myriad of possible analyses one can interpret from a study and the unaccounted influence of the data on that choice. Tackling this problem is fundamental to the long-term value of scientific research in every data-driven field, as highlighted by the frequent difficulty of reproducing scientific results.

One overly restrictive approach is to require researchers to pre-register their experiments, preventing them from accessing the dataset more than once. Over the past few years, a group of breakthrough papers from the computer science community addressed this question and showed that when the data is accessed only by sufficiently "stable" algorithms, then we can recalibrate the statistical tools to maintain accuracy despite the use of the data in both choosing an analysis and carrying it out. The first, and still best-known way to design such robust algorithms comes surprisingly from a security condition called "differential privacy". This condition limits the extraneous information that an analysis reveals about the data.

This work in computer science parallels efforts in statistics on "selective inference" and "false discovery rate". These similar areas of interest tackle different aspects of the same larger problem. Taken together, solutions to this problem hint at a tremendous opportunity; that of designing robust, reliable tools that allow data scientists to explore data and draw valuable insight from them, without mistaking characteristics of a particular data source for a general trend. Currently, we have access to more vastly varied data than ever before. Getting real value from these resources requires an interdisciplinary approach using ideas from statistics and areas of computer science, such as machine learning, data mining, and even data privacy, to design the tools we need.

## Data Dependent Algorithms

### *One size does not fit all — tailoring algorithms to data.*

The traditional approach to algorithm design is to focus on worst case scenarios, an excellent idea when mission critical tasks are in question. However, often the outcome is suboptimal in other scenarios. Imagine, for instance, preparing for a blizzard each time you leave your house. An exciting new direction in TCS is designing algorithms that perform better on "nice" data, the type of data that arises the most in practice. This new perspective aims at discovering unifying principles that can be used in a range of applications. The challenge is twofold: (1) characterize natural properties of data and define models that can be exploited for better performance. In particular, we seek data characteristics that persist across a broad range of computational tasks, (2) design algorithms that can exploit conducive features of data to improve their guarantees. Ultimately, the goal is to find data dependent algorithms that are as efficient as possible for each individual input.

This is a nascent line of research, but there are already some concrete examples where such an endeavor demonstrates promising success. There are two recent first steps: (1) the amazing developments in neural networks and deep learning are largely based on their power to represent massive collections of complex data (such as human face images, or MRI images) with very high accuracy. That leads us to believe we can develop better algorithms assuming that their input data is modeled in the same way. Bora et al. [1] recently developed highly efficient image acquisition methods (an underlying task for MRIs), for images modeled in this way, improving over the state-of-the-art by up to one order of magnitude, (2) an exciting new line of work by Kraska et al. [2] and by Mitzenmacher [3] characterizes "typical" data as "learnable" data. The study uses machine learning predictions about the input to improve the performance of various data indexing schemes that are widely used in software implementations, including standard programming languages and database systems.

## Algorithms from Algorithms

### *Desperately need an algorithm? Meta-algorithm to the rescue!*

An algorithm is a recipe to solve a problem. Designing algorithms is typically done by humans since it requires creativity and powers of formal reasoning. Given that computers are so powerful now, one wonders if they can supplement humans in the art and science of algorithm design. Can we create an algorithm that, given a problem description, designs an algorithm to solve it? Can we design an algorithm that self-improves over time, learning from its mistakes? Can an algorithm automatically find the optimal policy in decision making? Can an algorithm discover the best algorithm to solve a problem?

We have long known that such goals are mathematically impossible for arbitrary problems. But when considering real-world problems recent studies show these goals may be within reach. For a few problems, TCS researchers have shown how to make algorithms handle feedback and use the feedback to self-improve. The work lays out how to build algorithms that use problem-solution exemplars to design policies. In light of these early results, automated algorithm design is an area ripe for significant progress.

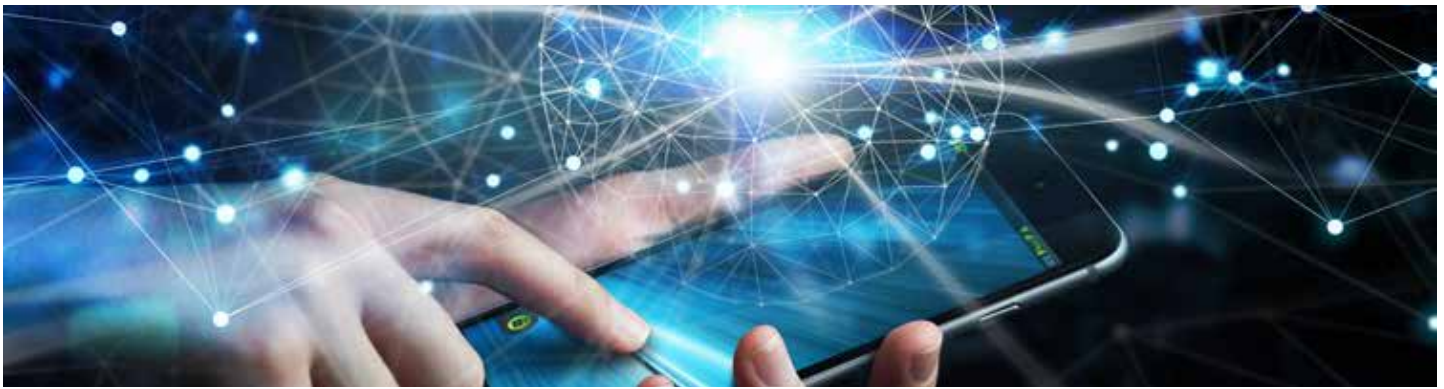

# FOUNDATIONS OF MACHINE LEARNING

It is undeniable that advances in theoretical computer science have had great positive impact on modern day machine learning as witnessed, for example, by the Kanellakis Award to Schapire for his work on "boosting" weak classifiers and to Cortes and Vapnik for their introduction of Support Vector Machines. The notion of "Probably Approximately Correct" learning and other notions at the core of computational learning theory were also developed by the TCS community, starting with the work of Leslie Valiant, then followed by many others.

Now with the central role machine learning plays in everyday life, new challenges and applications have emerged for which new foundations must be developed. For example, while traditional learning theory promotes the mantra of Occam's razor (i.e., learning 'simple' concept classes should be 'easier'), current techniques such as deep learning reject this conventional wisdom as they learn very complicated models with a large number of parameters. The ubiquity of machine learning also implies we need more robust learning algorithms that can perform well in the face of a small number of malignant users or corrupted training data, while simultaneously remaining efficient enough to fit on mobile devices such as cell phones. Machine learning systems should also maintain privacy and take game-theoretic considerations into account as users (people) interact with machine learning systems in strategic ways. All these considerations leave open a wide avenue for theoreticians to contribute to the development of the foundations of machine learning.

## Building Machine Learning Systems You Can Trust

### *How can we make machine learning systems provably safe and dependable?*

The fact that we currently trust machine learning with almost every aspect of our lives does not mean that our current machine learning toolkit is completely trustworthy. Existing machine learning systems can be easily manipulated to make mistakes which can compromise a number of critical applications. For example, ensuring that self-driving cars reliably detect pedestrians and obstacles, and making recommendation systems resilient to manipulations, are some of the challenges we have yet to tackle.

This quest critically requires developing a comprehensive algorithmic theory that enables us to precisely frame potential risks and develop principled approaches to mitigating them. One of the first strands of such a theory has already delivered a methodology for drawing correct conclusions from complex datasets in the presence of a significant degree of false data, leading to state-of-the-art empirical performance. A significant generalization and refinement of this theory is necessary to further improve practical performance. A major goal for TCS is to develop principled methodologies to address the broad set of challenges at the forefront of trustworthy machine learning.

## Foundations of Deep Learning

*A solid fundamental understanding of deep learning would allow us to overcome its drawbacks and broaden its applicability.*

Despite the huge impact that deep learning has had in practice, there are still many poorly understood aspects of this technology that are very different from classical methodology. Classical learning theory provides principled approaches for using data to choose simple prediction rules, whereas deep learning seems to achieve its outstanding performance through entirely different mechanisms: it relies on complex prediction rules, with an enormous number of adjustable parameters. Classical theory cannot explain why these complex prediction rules have outstanding performance in practice. In addition, it is also important to develop good mathematical models that explain data pertaining to language, images, and speech, and how their properties affect the complexity of learning. Gaining a theoretical understanding of these methods will be crucial for overcoming their drawbacks (see, for instance, the nugget on robustness), for developing algorithms that reduce the amount of trial and error involved in their deployment, and for extending them beyond the domains where they are currently applicable.

## Algorithms for Discovering Causal Structures in Data

*Uncovering causal relationships can reshape experimental design.*

One of the most important problems in analyzing data is detecting causal relationships among observable attributes. Can we distinguish correlation from causation? A well-studied abstraction for representing causality is a graphical model or network where attributes are linked to indicate the influence of one attribute over another. Some of the attributes may be latent or hidden from the observer. Their behavior is revealed only indirectly through the observed attributes. A major challenge for TCS is reconstructing plausible networks from a small set of observable behaviors.

For the fully observable case (no latent variables), efficient algorithms for learning the underlying network have only recently been obtained. These algorithms succeed even if the training set is small relative to the number of attributes. If we introduce a few latent variables, however, known solutions are slow to detect complex relationships among the observed attributes. The main goal is to break through this barrier and find efficient algorithms for the case of latent variables. Solutions will have a major impact in data mining and scientific modeling.

## Resource Aware Machine Learning

*Develop programs that can run on small devices and use as few resources as possible.*

Computing devices have become smaller (e.g., smart-phones, smart-watches) and more common in real-world environments (e.g., adaptive thermostats, digital voice assistants). Machine learning is an integral element and key driver of their success that has enabled new exciting experiences. Unfortunately, these devices have stringent resource constraints, including reduced processing capabilities, decreased power supplies, and bounded internet connections. Such constraints make deploying the typical, resource-intensive toolkit of machine learning impractical. Hence, it is challenging to create programs that are fast and reliable in this setting.

The question is how to design machine learning algorithms that can adhere to these resource constraints without compromising the offered functionality. TCS researchers have developed a mathematical abstraction of efficient programs that unifies the way we analyze various constraints. This allows us to better understand the fundamental tradeoffs among these real-world resources. Moreover, such an abstraction facilitates the development of machine learning methods that are aware of, and optimally adapt to, the available resources. These advances have helped reduce the energy consumption of standard computers by using a similar tradeoff analysis.

# Machine Learning for a Social and Strategic World

*Learning to design policies for people who have a vested interest in the outcome.*

When machine learning systems interact directly with people, they affect the way people behave. In recent years, corporations and government organizations have been steadily increasing their use of machine learning techniques to automate business, social, and economic decisions. For example, automated resume screening services are trained using past hiring and job performance data, and the calculation of insurance premiums is based on collective personal data. These consequential decisions shape people's lives and behavior which, in turn, shape the input to these critical decision-making systems.

Our goal is to design systems that can automatically uncover insights from these complex interactions, while avoiding major failures and harmful distortion of incentives and behavior. These methods also need to operate in unpredictable and ever-changing environments. It is therefore essential to develop methods that guarantee the performance of these systems even when the input changes. Adaptable machine learning systems that are designed in this way can shed light on how to improve and redesign existing social and economic policies to ensure reliable performance of the systems as well as the integrity of the societal forces they help to create. This is a vital step towards a world where technology serves to make society better, safe, and fair.

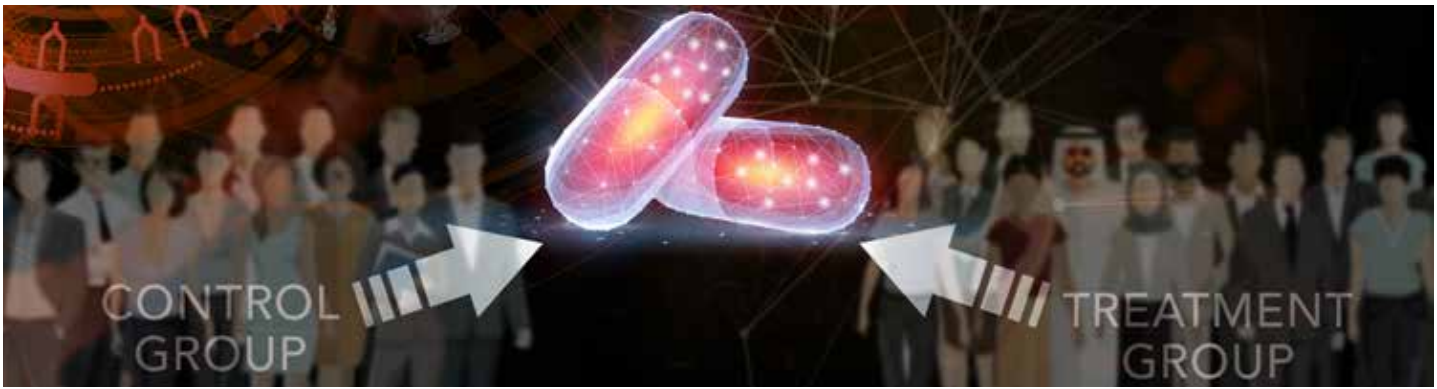

# OPTIMIZATION

## Linear System Solvers
*The essence and engine of scalable computation.*

Algorithm researchers strive to design better ways of solving problems that are central to many disciplines; one such fundamental problem is solving a set of linear equations in many variables. Systems of linear equations arise throughout science and engineering, from calculating stresses on truss structures, to fluid simulations, or to modeling the behaviors of electromagnetic waves. In many cases where linear systems don't exactly model the problem, they provide the steps that lead to the solutions. In fact, linear system solvers are the silent workhorses of much of large-scale computation today, including the search engines that power the modern internet.

Despite its storied history spanning centuries, we still do not know the fastest way to solve a system of equations. The approaches we learned to solve systems of equations in high school are too slow to solve the big problems we encounter in practice, even with the use of supercomputers. Over the last few decades, TCS researchers have developed entirely new techniques for solving equations in many variables that are much faster than classical methods. Many applications, (e.g., better integrated circuits, more fine-grained analysis of satellite images, and exploration of much larger social networks) have been greatly accelerated by the development of faster algorithms that solve the specific types of linear equations that arise within them.

For general linear systems, as well as many important subclasses, our best algorithms remain comparatively slow. Faster methods for solving systems of linear equations have led to, and will continue to lead to, accelerated drug design, better social network analytics, and more accurate recommendations of products for users. Advances in our understanding of this one fundamental problem impact all the areas in which it arises.

## Algorithms Improve Medical Trials.
*Making randomized control trials more efficient with algorithmic discrepancy theory.*

Randomized controlled trials (RCT) are the gold standard for evaluating medical treatments, and are a major tool of inquiry in science. Indeed, no novel treatment or drug can be approved without being vetted via an RCT, which around $500M is spent on annually.

The key to success in a randomized controlled trial is identifying two groups (the treated group and the baseline/placebo group) of test subjects whose profiles are sufficiently similar in terms of gender, age, health, and every other potentially relevant characteristic. The standard approach to achieving such similarity is to randomly partition the population of experimental subjects into the two groups. Such random assignment does lead to the desired similarity, but turns out to be suboptimal in terms of the most critical resource: the number of test subjects needed to draw sufficiently reliable conclusions. Indeed, by the 1970s, mathematicians working in discrepancy theory had proved that it is possible to divide people into two groups that are much more similar than the groups one gets from random assignment. However, these findings could not be used in practice because we did not know a practical way of finding these remarkable divisions.

This finally changed in 2010 with a breakthrough that launched the field of algorithmic discrepancy theory. Theoretical computer scientists working in this field are developing methods that enable us to efficiently compute divisions of people into groups whose makeups are shockingly similar. Through collaboration with statisticians these advances will create RCTs that provide us with more reliable conclusions while requiring fewer experimental subjects. This methodology is likely to accelerate the new drug and treatment development process, significantly reduce the cost of scientific studies across social sciences and medicine, and attain greater confidence in the identified findings.

# Cryptography

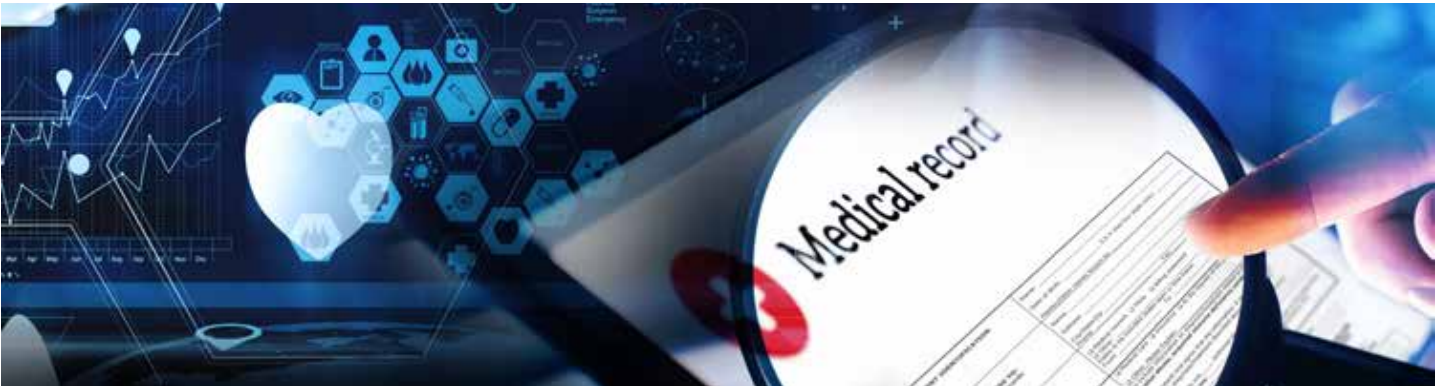

Cryptography has become the backbone of the Internet: it is deployed everywhere and secures our online banking, shopping, email, and other online activities. While traditionally cryptography was used to secure our communication, modern cryptography promises to secure even our computation. For example, Multi-Party Computation and Fully Homomorphic Encryption enables computation on encrypted data; and program obfuscation promises to hide secrets in our software. In the past decade, decentralized blockchains have gained traction, and there is a growing appetite for such rich cryptographic primitives to be deployed in the real world. The amazing developments from the cryptography community over the past ten years will lead to transformative new applications, in particular, it will allow computing and sharing of, and machine learning over, sensitive data while protecting our privacy.

## Can we Compute on Encrypted Data?

### Can you search the web without revealing your search query?

Can you run a python program without giving your input to the program? Can you contribute your data to a medical study while keeping the data private? Can we perform contact tracing and epidemic analysis without leaking users' private location traces?

Classical cryptographic systems are all-or-nothing; either you know the private key in which case you can see all of the data, or you don't, in which case you can't see any of the data and you cannot do much meaningful computation on the ciphertext. Fully-homomorphic encryption (FHE) enables running an arbitrary program on data while it remains encrypted, and provides a means to solve all these questions. Revolutionary developments in the last decade gave us fully homomorphic encryption schemes, harnessing beautiful mathematics from lattice-based cryptography. In fact, the rapid pace of research in the last decade accomplished great strides in making these tools practical. Several grand challenges remain, including improving the efficiency to be comparable to conventional cryptosystems such as RSA.

Concurrently, new threats to the security of our encryption systems continue to emerge. First, hardware attacks such as Spectre and Meltdown, aggravated by the remote storage of data, force us to harden our encryption systems against ever more powerful attacks. Secondly, the possibility of scalable quantum computers will render conventional encryption schemes, such as RSA, insecure. Can we construct *super-encryption schemes*, ones that not only allow us to compute on encrypted data, but also provide security against these more powerful attacks?

## Can we Hide Secrets in Software?

### Building obfuscation schemes that are provably secure and practically efficient.

Twelve years ago, the game-changing challenge of program obfuscation was suggested at this venue, we quote: "Progress in research on secure program obfuscation would lead to progress on important and long-standing open questions in cryptography with numerous applications."

The last seven years saw major breakthroughs, including the *first* feasibility result on program obfuscation. This presents the tantalizing possibility of masking not just data, but entire programs that can be used without anyone figuring out how they work, no matter the attacker's method. For the first time, this gives us hope that we can protect algorithms that are sent out into cyberspace.

Program obfuscation has turned out to be a swiss army knife for cryptography, presenting the possibility of building all sorts of dream cryptosystems. It enables encryption schemes with strong forms of access control, allowing a third party to learn functions of encrypted data, but not all of it. It allows us to securely patch and update software without revealing vulnerabilities, to prevent copyright infringement, to replace tamper-proof hardware — like Intel SGX — with software, and to secure programs for electronic commerce or national security applications that may be executed on insecure computers in the future.

Despite this exciting progress, existing obfuscators should primarily be viewed as feasibility results; bringing us closer to a dream that was previously believed to be unattainable. A lot more progress is needed before secure obfuscation becomes a practical reality. Moving forward, there are two grand challenges for obfuscation:

1.  Can we base the hardness of obfuscation on widely studied cryptographic assumptions? For example, building obfuscators that are provably unbreakable unless someone manages to factor large integers.

2.  How fast can obfuscators run? Existing algorithms will need to be made several orders of magnitude faster before they can be deployed in practice.

## Cryptography as Cartography: Mapping the Landscape of Computational Hardness

### *What can a computer not do? Can we use that for cryptography?*

The central paradigm in cryptography is to *use* computational hardness of central problems in computer science to construct secure systems. This puts us in a unique "win-win" situation; breaking the security of these systems would lead to interesting advances in other areas. For example, we can design encryption schemes that would lead to a breakthrough in number theory, the theory of error-correcting codes, or highly unexpected machine learning capabilities in the instance of a security violation.

Indeed, such encryption primitives are used in most of our interactions every day on the Internet. This being the case, developing a thorough understanding of computational hardness, both the kind we use today and that which we might be able to use tomorrow, is essential to make sure that our online banking systems and electronic commerce would still be secure even with new breakthroughs in machine learning and quantum computing.

In addition, this paradigm also lets us use cryptography to *explain* the difficulty of certain tasks in other areas of computer science. On the flip side, we can use cryptography to design concrete computational challenges and benchmarks that would serve as catalysts for progress in other areas such as quantum computing, machine learning algorithms, and verifying the correctness of programs. A prominent historical example is Shor's algorithm, which was motivated by the challenge of breaking the RSA encryption scheme and resulted in a huge interest in quantum computing.

# TCS for other domains

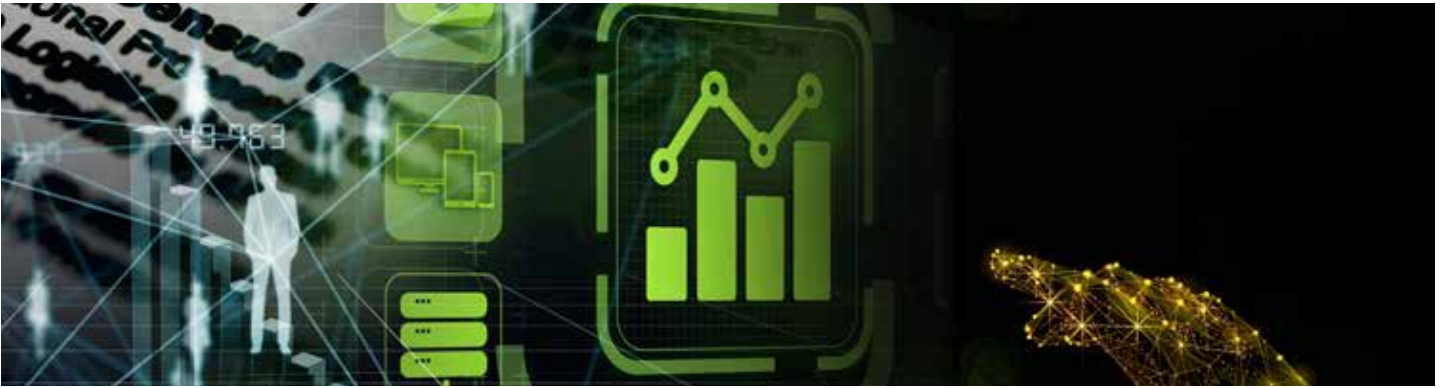

## A FIRM FOUNDATION FOR PRIVACY GROUNDED IN TCS

*Reliable methods for balancing privacy and utility, replacing ad hoc approaches.*

How can we reap the benefits of studying collections of large datasets without revealing sensitive information about the individuals to whom the data pertains to? This question is extremely challenging, as attackers have learned to evade current ad hoc defenses in increasingly devastating ways to extract sensitive data from seemingly anonymous information. Differential privacy came about in 2006 as a rigorous mathematical framework to identify what it would take for an algorithm to ensure the privacy of individuals against *any* attacker.

Over time, the TCS community has developed a strong understanding of how to reason about privacy and how to measure the accumulation of privacy risk as data is analyzed. We can now design very accurate algorithms that guarantee privacy for nearly every task from simple statistics to advanced ML techniques. Many players in industry and government are adopting this approach. The US Census Bureau is using these methods for the 2020 decennial census, and large companies such as Apple and Google are applying these techniques to ensure user privacy.

## Publishing Private Data for the Public Good

*How can we make data accessible for research and policymaking without compromising privacy?*

Large high-dimensional datasets containing personal information are commonly analyzed for multiple reasons. For example, the US Census Bureau's American Community Survey collects detailed information from a sample of American households. The results are used for major decisions about funding allocation and community planning.

In order to provide the widest accessibility to such data, we want to be able to publicly release a synthetic dataset — that is, an entirely new dataset that matches the statistical properties of the original data without corresponding to the real data of any individual. This is the approach used for the 2020 Census. Researchers have made great strides in developing synthetic data, but many challenges were also discovered.

It is impossible to generate a synthetic dataset that perfectly matches the real data. Decisions must be made about what kinds of analyses to prioritize and to what accuracy. From there stems the questions: Who gets to make these decisions? What technical tools are needed to ensure these decisions are aligned with society's goals? How can these decisions be communicated transparently? And how do we incorporate outside data sources and account for potential statistical biases?

Resolving these challenges would enable deeper analyses of private data than what is currently possible, as well as, reduce the disparate impact of privacy technologies on small subpopulations. In addition, solutions would catalyze advances in social sciences, genomics research, epidemiological modeling, and any other discipline where individual privacy is a major concern.

## The Privacy of AI

*How to build large-scale AI models while preserving your privacy.*

Many domains increasingly rely on large and complex AI models, but training these models requires massive datasets. In many cases, these required datasets contain personal and sensitive information: people's textual data, images and video content containing people and even health or financial data. Current AI models are known to memorize some of the training data and reproduce the information when prompted. The risk of these models leaking sensitive personal information from the training data is present and real. Moreover, the creation of the training datasets in itself creates a privacy risk, often making access to these valuable datasets restricted. For example, medical records distributed across organizations are currently silo-ed for privacy reasons, thus limiting the benefits derived from the integration of AI in these domains. Is it possible to mitigate privacy risks while realizing the full potential of AI technologies?

The answer to reducing these risks is privacy preserving technologies, such as differentially private and cryptographic tools, would allow the training algorithms to run without collecting data centrally and therefore guaranteeing that sensitive details cannot be extracted from the final model. Before implementing these tools, a couple of questions need to be answered: Can we develop algorithms that can be quickly trained to work on complex models and large distributed datasets? When is a loss in accuracy unavoidable to preserve privacy? And how will the use of these methods affect small subpopulations?

Successful resolutions to these questions would give us access to larger datasets to train AI models in a safe way that ensures privacy.

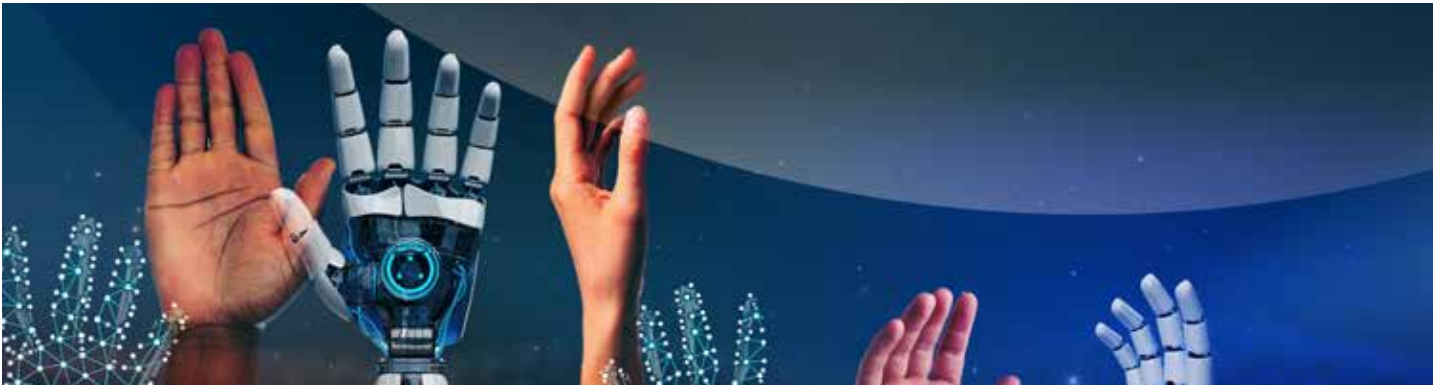

# TCS FOR SOCIAL GOOD

Algorithms are increasingly informing decisions deeply intertwined in our lives, from news article recommendations to criminal sentencing decisions and healthcare diagnostics. This progress, however, raises (and is impeded by) a host of concerns regarding the societal impact of computation. A prominent concern is that these algorithms need to be fair. Unfortunately, the hope that automated decision-making might be free of social biases is dashed due to the data with which the algorithms are trained and the choices made during their construction; left to their own devices, algorithms will propagate, or even amplify, existing biases in the data, the programmers, and the decisions of which features to incorporate and which measurements of "fitness" to apply. Addressing wrongful discrimination by algorithms is not only mandated by law and by ethics, but is essential to maintaining the public trust in the current computer-driven revolution.

Over the last decade, CS theory has revolutionized the landscape beliefs associated with algorithmic fairness. Many former notions of fairness mandate that the "average treatment" of a protected set of individuals (defined for example by gender or ethnicity) should be similar to that of the general population. For example, in the university admissions process statistical parity means that the fraction of accepted candidates from a protected group will be very close to the fraction of candidates from the general population. TCS researchers have shown in a sequence of studies that group fairness is easy to abuse; that natural group fairness definitions cannot be satisfied simultaneously in non-trivial scenarios; and that insisting on these definitions may even cause additional unexpected harm. While a single notion of fairness is impossible, theoreticians have introduced completely new families of definitions which allow for a more refined treatment of fairness. These notions can govern how individuals, or at least a multitude of groups, are treated. The new notions have already had substantial intellectual and practical impact.

## The Theory of Algorithmic Fairness

*Algorithms make decisions about us. We want these decisions to be fair and just.*

Society-facing automated decision systems have to perform a delicate balancing act. These systems need to be "good" at making decisions while simultaneously being fair and preventing discrimination against protected population subgroups. The theoretical computer science community has investigated these problems for nearly a decade.

Fairness is contextual; who should be treated similarly to whom depends on the task. One approach to algorithmic fairness separates the specification of fairness from the task of ensuring fairness is achieved. This places enormous burden on the specification, and allows a wide range of mathematical tools for achieving the task. The field has recently seen tantalizing socio-technical approaches to the specification problem.

By capturing unfairness, instead of fairness, the field can make progress as breaches in fairness are addressed. Here, however, theoretical investigation reveals an interesting phenomenon; several different types of statistical unfairness cannot be ruled out simultaneously. These kinds of mathematical limits focus research on articulating ambitious but achievable goals. This active area of research has produced gems with deep connections to the established fields of forecasting and regret minimization. Other investigations are more structural: Are systems composed of fair parts fair as a whole? Can properties of fairness be generalized? And is it better or worse, from a fairness perspective, to censor sensitive attributes in a dataset?

Many societal biases are systemic; as a result the ways in which individuals are presented to an algorithm can be problematic. The selection of variables, outcome proxies, and objective functions, do not treat similar individuals the same. The study of how to evaluate the risks, guarantees, and trade-offs associated with notions of fairness in algorithms is just beginning.

## Fair Resource Allocation

### *In a truly ethical society, how to be fair to all is the question.*

Whether it is distributing tax money, network access, school seats, satellites, defense resources or air traffic management, mechanisms that enable participants to come to a mutually agreeable allocation of resources are essential to the functionality of society. Fair division of resources has become a major field of study in computer science, mathematics, economics and political science. The domain seeks to answer whether provably fair outcomes exist in a wide variety of problems and how we can compute them. Computer science researchers have led the charge in making fair division possible at scale in environments with multitudes of participants, thereby bringing fair division to the masses. Important challenges in regards to even stronger notions of fairness still lie ahead: How can we divide goods that cannot be evenly divided in a manner that ensures no individual envies another's allocation? How can we define and impose fair division when the participants may lie about their preferences? And how can centralized allocation tools facilitate human interactions in a way that leads to fairer outcomes? The theory of fair resource distribution can provide allocations that are guaranteed to be fair in many applications of societal importance. For example, these methods are currently used for kidney exchanges, school seat assignments, dividing donations among food banks, inheritance division, divorce settlements, and splitting jointly purchased goods between friends.

## Auditing Algorithms

### *Telling fact from fiction in a world dominated by algorithms.*

Algorithms unleashed on a growing body of data are able to produce claims at an ever-increasing rate. For example, a machine learning algorithm applied to a medical record data set will be able to produce billions of credible-looking medical hypotheses. How can the scientific community independently verify these hypotheses?

We can describe how we want decision making systems to be fair, or nondiscriminatory, or avoid being biased. We can even design algorithms that satisfy all of these properties. But how do we make sure that systems we cannot directly inspect are trustworthy? We must have a way to *audit* systems that provide guarantees or certificates of trustworthiness.

To do this we have to address two challenges. First, the *power asymmetry:* the systems we want to audit are usually black boxes under the control of an untrusted third party with limited access for auditors. Second, the nature of the question to be asked: rather than trying to check if the system gives us correct answers, we want to find out whether the system is giving us *fair* or *unbiased* answers.

TCS is well placed to address the question of power: in areas ranging from interactive proofs to private information retrieval and the very architecture of distributed, private and secure computation, the framework of powerful demonstrators and weak verifiers show that we can solve surprisingly hard problems without full access to the system we are querying. The harder challenge lies in the distributional nature of the properties we wish to verify, whether it be a certificate of individual fairness, demographic parity, or other criteria that we wish to test for.

Diving deeper into this area of research will allow us to draw a better picture of what is possible and what is not. Such a direction will (and must) inform the larger legal and regulatory frameworks that are put in place for validating algorithmic decision systems.

## Algorithmic Foundations of Participatory Decision Making

### *Democracy needs more participation. Participation needs better algorithms.*

Although the world today is very different than it was a few decades ago, the practice of democracy hasn't fundamentally changed in centuries. To revitalize democracy, it is commonly believed that a higher degree of citizen participation is needed.

Theoretical computer science continues to play a major role in facilitating participation in a way that is representative and fair. In particular, local and national governments around the world are already eliciting and aggregating citizen preferences in order to select public projects to fund. TCS approaches seek to answer whether it is possible to compute outcomes that are guaranteed to be satisfactory to every possible group.

Similarly, randomly-selected citizens' panels that debate policy questions became commonplace in recent years. Issues of self-selection of volunteers and lack of fair representation on these panels can skew results and create bias threatening these democractic systems. The questions we must ask are: How can these panels be constructed in a way that is fair to volunteers, representative of the population, and transparent? And how can algorithms help panelists deliberate and reach a consensus? Theoretical computer science has the ability to ensure the fairness and representation of these systems and shape the future of democracy.

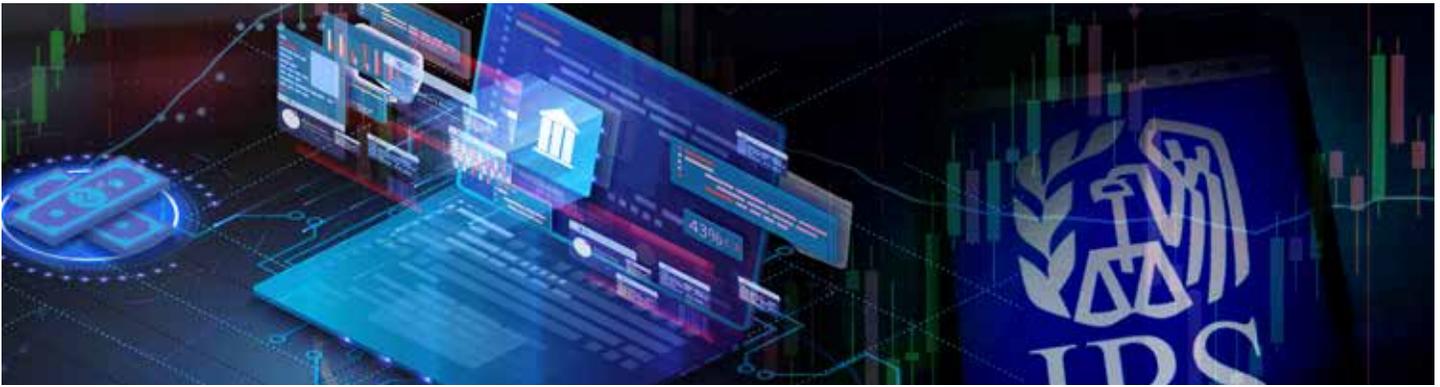

# AN ALGORITHMIC APPROACH TO ECONOMIC SYSTEMS

Algorithms and automated decision making have gradually permeated all levels of the U.S. economy. On the one hand, consider the daily decisions we make as consumers: How do you get to your next appointment – hail a taxi, share a ride, or rent a bike? What coffee machine should you buy? When you travel, where do you stay, where do you eat, and what do you do for fun? Increasingly, each of these decisions is made with both explicit and implicit direction from *platform marketplaces* run by firms such as Uber, Amazon, LinkedIn, Google, Facebook, and others. Individually, many of these marketplaces are large; in aggregate, their impact on the economy and on the shape of modern society is immense. On the other hand, institutions at the topmost levels of the economy have grappled with growing complexities by resorting to algorithmic approaches — from algorithmic trading in financial markets to large-scale Federal Communications Commission (FCC) spectrum auctions to barter markets for organ donation.

Traditionally, economic fields such as auction theory, contract theory, market design, and information design provided the principles that guide the configuration of economic markets and institutions. These foundations were recognized with Nobel Prizes in 2007 and 2020 for auction theory; in 2012 for market design; and in 2016 for contract theory. Yet, the theories they propose do not always scale well to the combinatorial complexities presented by modern markets, severely limiting their continued practical influence. TCS has stepped into this space to provide new economic theories built on a sound computational foundation.

## Platform Markets & Auctions at Global Scale

### *Who gets what and for how much — on the Internet?*

Auctions are not new, but fielding billions of interdependent auctions — run and monitored automatically — in complex environments is. TCS provides direct guidance on eliciting bids from participants along with pricing and allocating items. Online advertising, which is the main source of revenue for the big tech companies such as Google and Facebook, traces its lineage back to classical problems studied in TCS such as the online bipartite matching problem. Dynamic markets that match riders with drivers or bikes — that appear, disappear, and re-appear — at "NYC-scale" are guided toward efficient global solutions, all while making local decisions within seconds. In addition, applications such as Youtube and Spotify rely on core techniques in combinatorial optimization and learning theory to facilitate the fair representation of content creators in a diverse slate recommended to users. Indeed, TCS has even touched markets that operate without money, helping to build out barter markets for organs in the US and worldwide as well as matching markets for blood donation and charitable giving.

In the past decade, the TCS community made simplifying assumptions that allowed the field to make analytic progress, using these strong-but-simple models to help guide policy. The next step is understanding how to relax these assumptions to better capture the full reality of complex, decentralized, online marketplaces with strategic participants. What algorithms lead to good equilibria when all the participants are independently solving an online allocation problem for themselves? How does competition between different multi-sided platforms impact global social welfare, in both the short- and long-term? As platform markets continue to drive large parts of the online economy, these complexities will have growing implications for our society. Our push for the coming decade will be to provide a strong theoretical understanding of the dynamics of this complex environment.

## Optimization Drives the Sharing Economy
### *Everyone now runs their own bus company.*

While using a shared bike system to get around a city, we usually don't think about the planning and operational decisions that make the system run smoothly. For example, how do we ensure that there are enough bikes at each station for people to borrow and enough free slots so that people can also drop the bikes back off? The enormous number of users, and the uncertainty about when and where they want to pick up and drop off bikes, makes this an optimization problem of unprecedented scale. Similar challenges arise in delivering meals-on-wheels to needy families, or in matching ride-sharing cars to users: all these are systems that solve problems in real time, on-the-fly, and at huge scales.

These examples show how the sharing (digital) economy has given rise to challenging problems in resource management, logistics, and customer service. For many years TCS researchers have developed state-of-the-art methods to solve such problems, contributing to numerous success stories such as the AdWords systems. However, the diversity of these problems and the greater uncertainty means we need new formulations to model their practical difficulties. Research on the foundations of optimization algorithms will guide this work in two key ways. Firstly, it will provide abstractions to capture fundamental questions that span a range of applications within a common setting. Secondly, it will help answer these algorithmic questions by generating tools that guide the design, planning, and operational decisions. Both the underlying abstractions and the algorithmic tools can be used for other applications, some that are currently known, and many that are yet unanticipated.

## Economic Foundations for an Algorithmic World
### *Computer algorithms are evolving rapidly: Can economic institutions keep up?*

As computing becomes increasingly commodified, economic institutions are becoming increasingly algorithmic. Algorithmic trading is pervasive in financial markets. More and more people are finding work in a diverse range of fields through algorithmic platforms such as AirBnB, Uber/Lyft, and Mechanical Turk. Even within traditional firms, algorithmic hiring is a growing trend. Importantly, participants' behavior in these institutions is traditionally regulated (e.g., discriminatory hiring practices are illegal) in order to benefit society. When the participants are algorithms, however, regulation becomes problematic, or perhaps even intractable. One key example is algorithmic exploitation of arbitrage opportunities. Already, difficult-to-regulate algorithmic traders extract rents from financial markets by exploiting arbitrage opportunities before humans can even detect their existence. Imagine further algorithmic exploitation of tax loopholes, or algorithmic market manipulation in ride-sharing platforms — which are not far on the horizon.

On Page 29 we discussed approaches for auditing algorithms. Here we discuss an alternative approach that can supplement or supplant auditing — designing institutions to be resilient to or even benefit from participants' self-interested behavior. This approach fundamentally rethinks the design of economic institutions in two ways: first, modern markets should take advantage of (rather than collapse under the weight of) new state-of-the-art algorithms with heuristic, but not guaranteed, performance. Second, market regulation must have built-in robustness against algorithmic sophistication "growing with the state-of-the-art" built into them — otherwise, algorithmically sophisticated participants will be able to exploit other participants and the institution without providing any value in return.

These properties can be formalized through the TCS concept of *Price of Anarchy*, which measures the impact of strategic behavior on society's welfare within a system. Since its inception two decades ago, the Price of Anarchy has been a successful lens to study Internet routing protocols. Over the past decade, the paradigm has invaded other domains, such as auction design. The goal for the next decade is to influence an increasingly broader range of fields, and provide a rich theory to guide the design of economic institutions for strategic algorithmic participants.

# Workshop Participants

| Name | Affiliation |
| --- | --- |
| Scott Aaronson | University of Texas at Austin |
| Alex Andoni | Columbia University |
| Peter Bartlett | University of California Berkeley |
| Paul Beame | University of Washington Seattle |
| Mark Braverman | Princeton University |
| Mark Bun | Boston University |
| Mahdi Cheraghchi | University of Michigan Ann Arbor |
| Nikhil Devanur | Amazon |
| Ilias Diakonikolas | University of Wisconsin Madison |
| John Dickerson | University of Maryland |
| Cynthia Dwork | Harvard University |
| Anna Gilbert | Yale University |
| Seth Gilbert | National University of Singapore |
| Vipul Goyal | Carnegie Mellon University |
| Anupam Gupta | Carnegie Mellon University |
| Tom Gur | University of Warwick |
| Bernhard Haeupler | Carnegie Mellon University |
| Nika Haghtalab | Cornell University |
| Aram Harrow | Massachusetts Institute of Technology |
| Russell Impagliazzo | University of California San Diego |
| Piotr Indyk | Massachusetts Institute of Technology |
| Yuval Ishai | Technion - Israel Institute of Technology |
| Dakshita Khurana | University of Illinois Urbana-Champaign |
| Jonathan Kleinberg | Cornell University |
| Adam Klivans | University of Texas Austin |
| Gillat Kol | Princeton University |
| Ravi Kumar | Google |
| Shachar Lovett | University of California San Diego |
| Nancy Lynch | Massachusetts Institute of Technology |
| Aleksander Madry | Massachusetts Institute of Technology |
| Urmila Mahadev | California Institute of Technology |

| Name | Affiliation |
| --- | --- |
| Ruta Mehta | University of Illinois Urbana Champaign |
| Lorenzo Orecchia | University of Chicago |
| Rina Panigrahy | Google |
| Richard Peng | Georgia Institute of Technology |
| Seth Pettie | University of Michigan |
| Toniann Pitassi | University of Toronto |
| Yuval Rabani | Hebrew University of Jerusalem |
| Cyrus Rashtchian | University of California San Diego |
| Ilya Razenshteyn | Microsoft Research Redmond |
| Omer Reingold | Stanford University |
| Guy Rothblum | Weizmann Institute of Science |
| Aviad Rubinstein | Stanford University |
| Grant Schoenebeck | University of Michigan |
| C. Seshadhri | University of California Santa Cruz |
| Elaine Shi | Carnegie Mellon University |
| David Shmoys | Cornell University |
| Adam Smith | Boston University |
| Daniel Spielman | Yale University |
| Thomas Steinke | Google |
| Kunal Talwar | Apple |
| Li-Yang Tan | Stanford University |
| Eva Tardos | Cornell University |
| Jonathan Ullman | Northeastern University |
| Vinod Vaikuntanathan | Massachusetts Institute of Technology |
| Sergei Vassilvitskii | Google |
| Prashant Vasudevan | University of California Berkeley |
| Vijay Vazirani | University of California Irvine |
| Santosh Vempala | Georgia Institute of Technology |
| Suresh Venkatasubramanian | University of Utah |
| Matt Weinberg | Princeton University |
| Ryan Williams | Massachusetts Institute of Technology |
| John Wright | University of Texas at Austin |
| Henry Yuen | Columbia University |